\documentclass[a4paper,11pt]{article}
\usepackage{jheppub}
\usepackage{amsmath}
\usepackage{amsfonts}
\usepackage{amssymb}
\usepackage{textcomp}
\usepackage{graphicx}
\usepackage{bm}
\usepackage{color}
\usepackage{verbatim}
\usepackage{subfig}
\usepackage{graphicx,color}
\usepackage{epsfig}

\usepackage{amsfonts,amsmath,amssymb}

\newcommand{\p}{\partial}

\newcommand{\rar}{\rightarrow}

\newcommand{\qf}{ \delta Q_2}
\newcommand{\af}{ \delta A_2}

\begin{document}

 \title{Commensurability effects \\ in holographic homogeneous lattices}
 
 \author[a]{Tomas Andrade}
 
  \author[b]{Alexander Krikun}
 
 \affiliation[a]{Rudolf Peierls Centre for Theoretical Physics \\ University of Oxford, 1 Keble Road, Oxford OX1 3NP, UK}

\affiliation[b]{Instituut-Lorentz \footnote{since 01 Oct 2015}, Universiteit Leiden \\ P.O. Box 9506, 2300 RA Leiden, The Netherlands} 
 
\affiliation[b]{NORDITA \footnote{before 25 Sep 2015},
KTH Royal Institute of Technology and Stockholm University \\
Roslagstullsbacken 23, SE-106 91 Stockholm, Sweden. }

\affiliation[b]{On leave from Institute for Theoretical and Experimental Physics (ITEP), 
B. Cheryomushkinskaya 25, 117218 Moscow, Russia }

\emailAdd{tomas.andrade@physics.ox.ac.uk}
\emailAdd{krikun@lorentz.leidenuniv.nl}

\abstract{An interesting application of the gauge/gravity duality to condensed matter physics is the description of a 
lattice via breaking translational invariance on the gravity side. By making use of global symmetries, it is possible to do 
so without scarifying homogeneity of the pertinent bulk solutions, which we thus term as ``homogeneous holographic lattices." 
Due to their technical simplicity, these configurations have received a great deal of attention in the last few years and have been shown to correctly describe momentum relaxation and hence (finite) DC conductivities. 

However, it is not clear whether they are able to capture other lattice effects which are of interest in condensed matter. In this 
paper we investigate this question focusing our attention on the phenomenon of commensurability, which arises when the 
lattice scale is tuned to be equal to (an integer multiple of) another momentum scale in the system. We do so by studying 
the formation of spatially modulated phases in various models of homogeneous holographic lattices. 

Our results indicate that the onset of the instability is controlled by the near horizon geometry, which for insulating solutions does carry information 
about the lattice. However, we observe no sharp connection between the characteristic momentum of the broken phase and the lattice pitch,
which calls into question the applicability of these models to the physics of commensurability.
}

\maketitle

\section{Introduction}

A considerable contribution to the recent development of holographic models of condensed matter comes from the description 
of an ionic lattice. The lattice plays a fundamental role in all materials, giving rise to phenomena such as rotation and translation symmetry breaking, formation of the band structure, finite resistivity etc. It is understood that the translational symmetry breaking
caused by the lattice is necessary to obtain finite resistivity in holographic models and to study metallic, insulating and superconducting phases as well as the transitions between them 
\cite{Donos:2012js, Donos:2013eha, Andrade:2013gsa, Donos:2014oha, Donos:2014uba, Gouteraux:2014hca, Taylor:2014tka, Horowitz:2012ky, Horowitz:2013jaa, Donos:2014yya, Rangamani:2015hka, Erdmenger:2015qqa, Andrade:2014xca, Kim:2015dna, Ling:2014laa, Ling:2014saa}. 
Generically, quite involved numerical calculations are required in order to introduce even the simplest one-dimensional periodic lattice in the gravitational model \cite{Horowitz:2012ky} (see also \cite{Donos:2014yya, Rangamani:2015hka}), however, several setups have been proposed which simplify the treatment considerably. By making use of certain global symmetries of the bulk theories, the so called ``homogeneous lattice'' models are built in such a way that the generic set of partial differential equations of motion resulting from breaking translational symmetry along the boundary directions  is reduced to a set of ordinary differential equations in the holographic coordinate. 
Examples of this kind include the helical lattice model \cite{Donos:2012js}, the Q-lattice model \cite{Donos:2013eha} and 
the linear axion model \cite{Andrade:2013gsa}. These homogeneous lattices have proven to be quite successful 
in describing translation symmetry breaking and the associated finite resistivity in holographic models of metals by reproducing 
the expected Drude physics\footnote{Breaking diffeomorphism invariance by adding a mass to the graviton has also been considered
as an alternative holographic model for momentum relaxation \cite{Vegh:2013sk}.}. 
Moreover, metal-insulator transitions
have been studied in \cite{Donos:2013eha, Donos:2012js, Ling:2015dma}, and ``strange metals'' have been obtained \cite{Hartnoll:2015sea, Lucas:2014sba}. 
Being substantially simpler at the technical level than their inhomogeneous counterparts, the homogeneous lattices provide an attractive 
setup for various theoretical studies. Nevertheless, the question of to what extent one may make use of them as 
models for ionic lattices remains open.

Apart from relaxing momentum, the other important role of the lattice is to provide a certain length scale, i.e. the period of the background structure. This length scale plays an important role in various commensurability effects. The essence of the phenomenon is that 
interactions may lead to the spontaneous formation of inhomogeneous superstructures in the electronic subsystem 
with a certain wavelength, which may or may not be commensurate with the lattice spacing. 
A simple mechanical analogy to such phenomena is provided by the Frenkel-Kontorova model \cite{FKbook}, which describes a
one-dimensional array of masses and springs in the presence of an external periodic potential. 
When the two structures are commensurate, i.e. when their periods are equal or one is an integer multiple of the other, characteristic phenomena appear such as the Mott insulator state.
Incommensurability of the structures leads to moire\'{e}-type effects, which can be seen as the formation of charge density waves.  
The aim of this paper is to examine whether similar commensurability effects can be captured by the holographic homogeneous lattice models.
 
In the Frenkel-Kontorova model the preferred periodicity of the masses and springs is defined by their interactions, i.e. the superstructure is formed spontaneously.  In the holographic context, the spontaneous formation of spatial superstructures has been extensively studied. 
Examples involve the spontaneous formation of helices \cite{Nakamura:2009tf, Ooguri:2010kt, Donos:2012wi}, charge density waves \cite{Donos:2013gda}, helical superconductors \cite{Donos:2011ff} and various types of striped orders \cite{Jokela:2014dba, Withers:2014sja, Krikun:2015tga, Donos:2013woa, Erdmenger:2013zaa}. 
In this work we shall use the spontaneous formation of the helical structure as a way to probe the holographic 
homogeneous lattice, concentrating on the study of the pitch of the helical instability and the critical temperature 
at which it occurs. 

We will study several homogeneous lattice backgrounds as a substrate. These are, 
in increasing order of complexity, the linear axion \cite{Andrade:2013gsa}, the Q-lattice \cite{Donos:2013eha} and the Bianchi VII helix \cite{Donos:2012js}. 
We avoid any explicit couplings in the Lagrangian between the field that characterizes the formation of the instabilities 
and the matter sources of the geometry. This allows us to concentrate on the study of the gravity sector of the model which is responsible for the description 
of the holographic lattice\footnote{Because we allow the background configuration to consistently backreact on the modes 
that drive the instabilities, the unperturbed values of the matter fields also enter in the mode equations of motion.}.
Based on this, it is in principle unclear whether one can regard the linear axion and the Q-lattice as periodic substrates because,
even though they both possess a characteristic momentum scale, their line elements are translational invariant. Therefore, our most intriguing 
example is the helical background, since in this case the geometry explicitly depends on the spatial coordinates in a periodic way. 
Moreover, our probe is exactly of the same structure as the helical source which gives rise to the geometry, hence it is reasonable to expect a nontrivial interplay between the substrate and the probe. 

If present, the commensurability effects will manifest themselves as a certain feature happening as we dial the momentum scale of 
the underlying background, $k$, to coincide with the natural momentum of the probe helix -- the momentum which the 
helix assumes in the translational invariant Reissner-Nordstr\"om background, $p_c^{RN}$. Because the lattice is not driving 
the helix away from its natural period, we expect this configuration to have lower free energy and higher transition temperature than 
in those for which $k \neq  p_c^{RN}$. 

Another interesting possibility which we would interpret as the singling out of the lattice momentum scale would be the 
interference between the helix and the background.
More concretely, one could expect that, if the period of the spontaneously arising structure is tuned to be equal to the period 
of the lattice, these two periodic structures would interfere which should also have some detectable effect on the free energy and the critical 
temperature of the phase transition. 

The paper is organized as follows. In Section \ref{sec:flat} we review the spontaneous formation of the helical condensate in the
translational invariant case of Reissner-Nordstr\"om, which will be useful for later comparison and to set our notations and methods. 
In Sections \ref{sec:axion}, \ref{sec:Qlattice} and \ref{sec:helix} we study the helical instabilities in the linear axion, Q-lattice and helical backgrounds, respectively. We conclude in Section \ref{Conclusion}. 
We write down the equations of motion in Appendix \ref{app:eoms}, provide the details of our numerics in Appendix \ref{app:relax} and in Appendix \ref{app:sigmaDC} we present a derivation of the formula for DC conductivity in the helical background.

\section{\label{sec:flat}Spatially modulated instabilities in Reissner-Nordstr\"om}

Let us begin with the discussion of the formation of spatially modulated phases in the simplest example of a 
translational invariant background \cite{Nakamura:2009tf}. 
%
%
To this end, \cite{Nakamura:2009tf} considered a five dimensional bulk model described by the  Lagrangian
\begin{equation}\label{spont_action}
 \mathcal{L} =  \mathcal{L}_0 +  \mathcal{L}_{CS}
 \end{equation} 
\noindent where 
\begin{align}\label{LRN}
	&\mathcal{L}_0 = \sqrt{-g} \left( R + 12 -\frac{1}{4}  F_{IJ}F^{IJ} \right), \\
\label{LCS}
	&\mathcal{L}_{CS} = \frac{\gamma}{3!} \epsilon^{IJKLM}A_I F_{JK}F_{LM}.
\end{align}

Here and henceforth the AdS curvature radius is set to one. 
Capital Latin indices correspond to space-time quantities, 
$F_{IJ}$ is the field-strength associated to an abelian $U(1)$ form $F_{IJ} = \partial_I A_J - \partial_J A_I$ and $\epsilon^{IJKLM}$
is the Levi-Civita symbol which takes the values ${0, \pm 1}$. The coupling $\gamma$, which controls the strength of the 
Chern-Simons (CS) term in \eqref{spont_action}, is dimensionless. The presence of the CS term implies that the dual current
associated to the $U(1)$ field is anomalous.

\begin{figure}[ht]
\centering
 \includegraphics[width=0.6\linewidth]{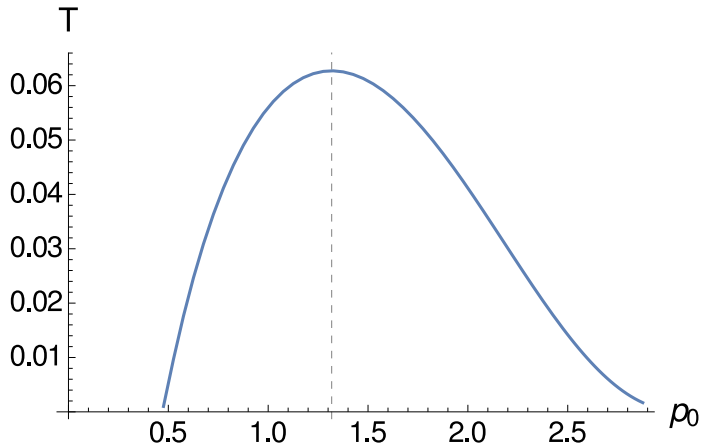}
 \caption{\label{bell_RN} The bell curve in the RN background for $\gamma$=1.7. The dashed vertical line 
 corresponds to $p_0 = p_c^{RN}$.}
\end{figure}

The equations of motion admit as a solution the Reissner-Nordstr\"om AdS (RN) black hole
 \begin{align}
 \label{metricRN}
 ds^2 &= -U(r) dt^2 + \frac{dr^2}{U(r)} + r^2 (dx^2 + dy^2 + dz^2),\\
 U(r) &= r^2 - \left(r_h^2 + \frac{\mu^2}{3} \right)\frac{r_h^2}{r^2} + \frac{\mu^2}{3} \frac{r_h^4}{r^4}, \\
  A &= \mu \left(1 - \frac{r_h^2}{r^2}\right) dt,
\end{align}
\noindent where $\mu$ is the chemical potential, $r_h$ is the radius of the horizon, defined as the largest root of $U(r_h)=0$, 
and the temperature is fixed by the surface gravity at the horizon
\begin{align}
\label{temperature}
T = \frac{U'(r_h)}{4 \pi}.
\end{align}
The main result of \cite{Nakamura:2009tf}  is that this normal state solution becomes unstable against spatially dependent 
perturbations as we lower the temperature, provided the CS coupling is large enough. 
To make our discussion self-contained, we summarize here the main ingredients of the calculation following the notation and 
conventions of \cite{Donos:2012wi}. 
In order to see the formation of the spatially modulated phase, one studies the spectrum of fluctuations around the 
background solution \eqref{metricRN}. More concretely, 
the presence of a nontrivial mode with zero frequency and non-zero momentum in the spectrum would signal the onset 
of the instability. 
Such unstable modes can be captured considering the ansatz for the fluctuation of the spatial components of the gauge field 
\begin{equation}
\label{spont_ansatz}
 \delta A = \af (r) \, \omega_2^{(p)},
\end{equation}
\noindent where $\omega_2^{(p)}$ belongs to the set of helical 1-forms with pitch $p$
\begin{align}
\label{helical_forms}
\omega^{(p)}_1 &= dx, \\ 
\notag
\omega^{(p)}_2 &= \cos(p x) d y - \sin(p x) dz, \\
\notag
\omega^{(p)}_3 &= \sin(p x) d y + \cos(p x) dz. 
\end{align}
%
%
Due to the fact that the $t$-component of $A$ has a nonzero background value, the profile $\delta A$ couples to the fluctuations of metric, 
which can be written in terms of the helical forms \eqref{helical_forms} as
\begin{equation}\label{delta ds2 RN} 
	\delta ( ds^2 ) = 2  \qf (r) \, dt \, \omega_2^{(p)}
\end{equation}
The fluctuation equations are given in \eqref{lin eom axion 1} and \eqref{lin eom axion 2}.  
The search for unstable modes entails finding a nontrivial solution to the set of coupled 
ODE's for $\af$, and $\qf$. The solutions to these differential equations are 
fully determined once we specify boundary conditions. We demand regularity at the horizon, which implies
that near $r = r_h$ the fields behave as
\begin{align}\label{bc H RN}
r\rar r_h: \qquad	\af(r) &= A_2^{(h)} + \ldots. \qquad  \qf(r) = Q_2^{(h)} (r - r_h)  + \ldots,
\end{align}
\noindent where $A_2^{(h)}$ and $Q_2^{(h)}$ are arbitrary constants and the ellipses 
denote subleading analytic terms in $(r- r_h)$. Since we are interested in the {\it spontaneous}
breaking of translational symmetry, we impose boundary conditions in the UV which correspond to setting to zero 
the sources of the corresponding dual operators, i.e. we demand that near $r \to \infty$ the perturbations behave as
\begin{align}\label{bc dM RN} 
r \rar \infty: \qquad	\af(r) &= \frac{ A_2^{(2)} }{r^2}  + \ldots, \qquad	\qf(r) = \frac{ Q_2^{(2)}  }{r^2}  + \ldots,
\end{align}
\noindent where $A_2^{(2)} $ and $Q_2^{(2)}$ are arbitrary constants and the ellipses denote 
subleading analytic terms in $1/r$.
For a given temperature, the unstable modes (if any) are found at specific values of $p=p_0(T)$, which form a characteristic 
``bell curve, '' see Fig.\,\ref{bell_RN}. The solutions to this Sturm-Liouville problem are obtained numerically. In this example and in other backgrounds we use a shooting method which takes advantage of the linearity of the problem, see e.g. \cite{Krikun:2013iha} or Sec.4.3.2 in \cite{Silaev}

The tip of the bell curve corresponds to the unstable mode which is observed at the largest 
temperature $T_{max}$. Thus, this is the first mode which develops upon cooling of the system and 
the corresponding value of the 
critical pitch $p_c$ determines the periodicity of the resulting new state which forms. 
As in \cite{Nakamura:2009tf}, we fix $\gamma=1.7$ throughout this paper. For this value of the CS coupling, 
in the RN background the tip of the bell curve is characterized by
\begin{equation}
\label{RNvalues}
p_c^{RN} = 1.31, \qquad T_{max}^{RN} = 0.0627.
\end{equation}
Here and in what follows we express all numerical dimensionful quantities in units of the chemical potential $\mu$. 
The quantity $p_c^{RN}$ is the natural momentum of the free helix, this is, in the absence of momentum relaxation.
As mentioned in the Introduction, it will play an important role in the following analysis of commensurability effects.

It is worth emphasizing that, as explained in \cite{Nakamura:2009tf}, the development of these instabilities can be 
understood as an IR effect. The argument is as follows\footnote{A similar line of reasoning can be used to explain 
the appearance of the superconducting instability of AdS black holes \cite{Hartnoll:2008kx}.}: at very small temperatures, the 
near horizon geometry is approximately $AdS_2 \times R^3$, so that the fluctuation equations reduce to field equations in $AdS_2$ with 
effective masses that depend on the spatial (boundary) momentum $p$ of the given modes. For certain values of $p$, the effective
near horizon masses 
violate the $AdS_2$ Breitenlohner-Freedman bound \cite{Breitenlohner:1982jf,Breitenlohner:1982bm} which in turn indicates that the full geometry should be unstable. 
In the remaining sections we will find evidence that the helical instabilities continue to be essentially an IR phenomenon after 
we include momentum relaxation. 

Strictly speaking, the new ground state cannot be obtained by a perturbative analysis \cite{Ooguri:2010kt}. The full nonlinear 
problem was solved in \cite{Donos:2012wi} and the new phase was found by minimization of the thermodynamic potential. 
It was shown that the phase transition is of second order and that at temperatures close to $T_{max}$ the pitch 
of the spontaneously formed helical state is indeed close to $p_c$. 

In what follows we will restrict ourselves to the type of perturbative analysis presented in this section. This will allow us
to determine the onset of the instability which gives rise to the spontaneous formation of a helical state in the homogeneous 
lattices under consideration.

\section{\label{sec:axion}Linear axion model}

Our first example of the formation of a helical phase on a background with broken translational symmetry is given 
by the linear axion model of \cite{Andrade:2013gsa} coupled to a CS term. We consider the bulk Lagrangian
\begin{equation}
	\mathcal{L} = \mathcal{L}_0 + \mathcal{L}_{{\rm axions}} + \mathcal{L}_{CS}
\end{equation}
\noindent where $\mathcal{L}_0$ and $\mathcal{L}_{CS}$ are given by \eqref{LRN} and \eqref{LCS} respectively, and
\begin{equation}
	\mathcal{L}_{{\rm axions}} = - \frac{\sqrt{- g}}{2} \sum_{a = 1}^3 (\partial \psi_a)^2  
\end{equation}
Here $\psi_a$ are three real, massless scalar fields. 
The following background is an exact solution of the equations of motion\footnote{This solution was previously 
found in \cite{Bardoux:2012aw} in a different context.}
\begin{align}
\label{metric linear axion}
 ds^2 &= -U(r) dt^2 + \frac{dr^2}{U(r)} + r^2 \delta_{ij} dx^i dx^j,\\
 U(r) &= r^2 - \frac{\alpha^2}{4} - \left(r_h^2 + \frac{\mu^2}{3} \right)\frac{r_h^2}{r^2} + \frac{\mu^2}{3} \frac{r_h^4}{r^4}, \\
  A &= \mu \left(1 - \frac{r_h^2}{r^2}\right) dt, \qquad \psi_a = \alpha \delta_{a i}  x^i 
\end{align}
\noindent where $i, j$ are indices that run over the boundary spatial dimensions, e.g. $x^i = (x,y,z)$. This solution 
can be characterized by two dimensionless parameters which we take to be $T/\mu$ and $\alpha/\mu$.
Note that, despite the fact that the geometry is isotropic and homogeneous, the background manifestly breaks translational 
symmetry through the dependence of the $\psi_a$ on the boundary coordinates. 
As a consequence, the delta-function at zero frequency in the optical conductivity which occurs in the RN solution is resolved 
\cite{Andrade:2013gsa}. Interestingly, as pointed 
out in \cite{Kim:2014bza}, the resulting peak is in fact Drude for small values of $\alpha/T$,  but deviations from Drude 
physics are observed as we increase $\alpha/T$. Similar ``coherent/incoherent" transitions have also been studied in 
the context of holography in \cite{Davison:2014lua, Davison:2015bea}.  
For any value of $\alpha$, the DC conductivity can be evaluated to be 
\begin{equation}
 	\sigma_{DC} = r_h \left( 1 + \frac{4 \mu^2}{\alpha^2} \right)
\end{equation} 
\noindent where $r_h$ should be understood as a function of $T$, $\alpha$ and $\mu$.  At zero temperature, $\sigma_{DC} $ takes the 
value
\begin{equation}\label{sDC axion T0}
 	\frac{\sigma_{DC}}{\mu} \bigg|_{T = 0} = \frac{1}{2 \sqrt{6}} \left( 1 + \frac{4 \mu^2}{\alpha^2} \right) \sqrt{4+3 (\alpha/\mu)^2} 	
\end{equation} 

\begin{figure}[ht]
\centering
 \includegraphics[width=1\linewidth]{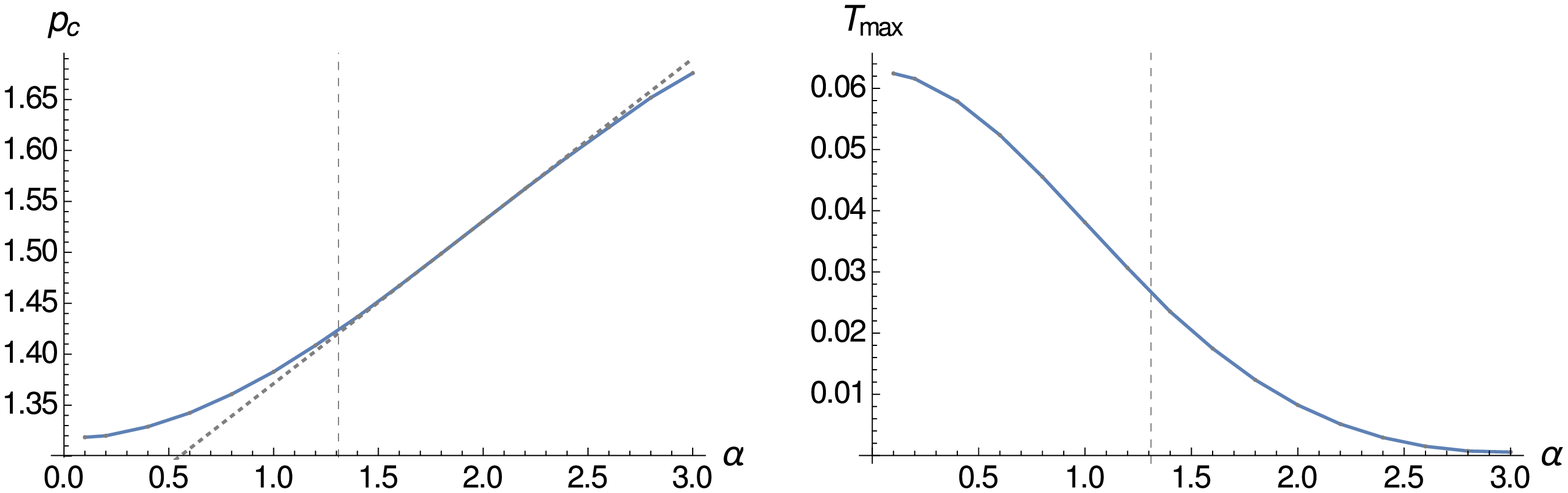}
  \caption{\label{axion_results} The position of the top of the bell curves in linear axion background.}
The vertical grid lines show the natural momentum scale of the free helix $p_c^{RN}$.
\end{figure}

It turns out that the minimal set of perturbations
we need to study in order to detect the onset of the spatially modulated instabilities is the same as in the translationally 
invariant case, i.e. \eqref{spont_ansatz}, \eqref{delta ds2 RN}. The resulting equations of motion are 
\eqref{lin eom axion 1}, \eqref{lin eom axion 2} and the boundary conditions are given by \eqref{bc H RN}, \eqref{bc dM RN}. 
In this case, however, we have a one dimensional parameter space of backgrounds to explore, which we parametrize by $\alpha/\mu$. 
Therefore, we numerically study the unstable modes on the plane $(p, T)$ as described in Section \ref{sec:flat} at fixed 
$\alpha/\mu$. We find that at all values of $\alpha/\mu$ the curves on the $(p, T)$ plane are qualitatively just like the bell in Fig.
\ref{bell_RN}, but with peaks of varying location $p_c$ and height $T_{max}$. We summarize our results in Fig. \ref{axion_results}. \\   

We find no salient feature when the background momentum scale $\alpha$ becomes equal to the free spontaneous 
momentum $p_c^{RN}$ (the vertical, dashed lines on the each panel of Fig. \ref{axion_results}), which we interpret
as the absence of commensurability effects. 
Nevertheless, we do observe a nontrivial interplay between the lattice and the probe. One clearly sees that $p_c$ is affected by the background and grows linearly with $\alpha$ for large $\alpha$ (the linear growth is approximated by $p_c = 1.2117 + 0.1595 \alpha$), while $T_{max}$ decays rapidly.

It is worth commenting that the properties of the instabilities in this background are similar to those of RN only for very small $\alpha/\mu$. 
In addition, notice that, because the $\psi_a$ are independent of $r$, the operators that control the lattice are marginal in the IR. 
Therefore, for any value of $\alpha/\mu \sim O(1)$ the near horizon geometry departs considerably from RN\footnote{As we shall discuss in 
detail later on, this is consistent with the fact that $\sigma_{DC}$ in \eqref{sDC axion T0} is finite as we approach $T=0$.}. 
This is an indication that the instability that drives the system towards the spatially modulated phase is an IR effect, just as
in the translational invariant case.

\section{\label{sec:Qlattice}Q-lattice}

In this section we consider the instability towards the formation of the helix on a 5-dimensional generalization 
of the Q-lattice geometry of \cite{Donos:2013eha}. More concretely, we focus on the Lagrangian 
\begin{equation}
	\mathcal{L} = \mathcal{L}_0 + \mathcal{L}_{\rm Q-lattice} + \mathcal{L}_{CS}, 
\end{equation}
\noindent with $\mathcal{L}_0$, $\mathcal{L}_{CS}$ in \eqref{LRN},\eqref{LCS} and
\begin{equation}
	\mathcal{L}_{{\rm Q-lattice}}= - \sqrt{- g} ( |\partial \phi|^2 + m^2_\phi |\phi|^2  ), 
\end{equation}
\noindent where $\phi$ is a neutral, complex scalar field. In close analogy with \cite{Donos:2013eha}, 
we consider the ansatz
\begin{gather}
\label{metric Q-lattice}
 ds^2 = -U(r) dt^2 + \frac{dr^2}{U(r)} + e^{2 v_1(r)} dx^2 + e^{2 v_2(r)} (dy^2 + dz^2),\\
 \notag
  A = a(r) dt, \qquad \phi = e^{i k x} \chi(r).   
\end{gather}
Note that the geometry is isotropic but inhomogeneous because the $x$ coordinate is singled out. Because of the global $U(1)$
symmetry associated to the complex scalar, the $x$-dependence drops out of the equations of motion so that the resulting 
equations for the unknowns $\{U(r), v_1(r), v_2(r), a(r), \chi(r)\}$ are ODE's in the radial coordinate $r$. 
They are given by \eqref{bck eom qlattice 1}-\eqref{bck eom qlattice 5}.

For concreteness, we choose the mass of the scalar field to be $m^2 = - 15/4$ since this simplifies the boundary asymptotics. 
With this choice, the UV expansions can be written as
\begin{align}\label{bcs qlattice back}
r\rar \infty: \qquad	U(r) &= r^2 + \ldots, \qquad  v_i(r) = \log r + \ldots,\\
\notag
	a(r) &= \mu + \ldots , \qquad \chi(r) = \frac{1}{r^{3/2}} ( \lambda + \ldots)
\end{align}
\noindent The subleading terms consists of regular powers of $1/r$ (with the global factor of $r^{-3/2} $ in $\chi$). 
We assume that there is a regular finite temperature horizon at $r = r_h$. Near this point, the asymptotics can be written as
\begin{align}
r\rar r_h: \qquad	U(r) &= (r-r_h) U_h + \ldots, \qquad e^{2 v_i(r)} =  e^{2 v_{i h}} + \ldots, \\
\notag
	a(r) &= (r-r_h) E_h + \ldots, \qquad 	\chi(r) = \chi_h + \ldots,
\end{align}
\noindent where the subleading terms can be written as regular power series in $(r-r_h)$. As explained in 
\cite{Donos:2013eha},  we expect to find a 3 parameter family of solutions, which we choose to be the scale invariant 
combinations $T/\mu$, $\lambda/\mu^{3/2}$ and $k/\mu$.  

\begin{figure}[ht]
\subfloat[][\label{Q_pc_3D} Dependence of $p_c$ on the background parameters.]{
  \includegraphics[width=0.3\linewidth]{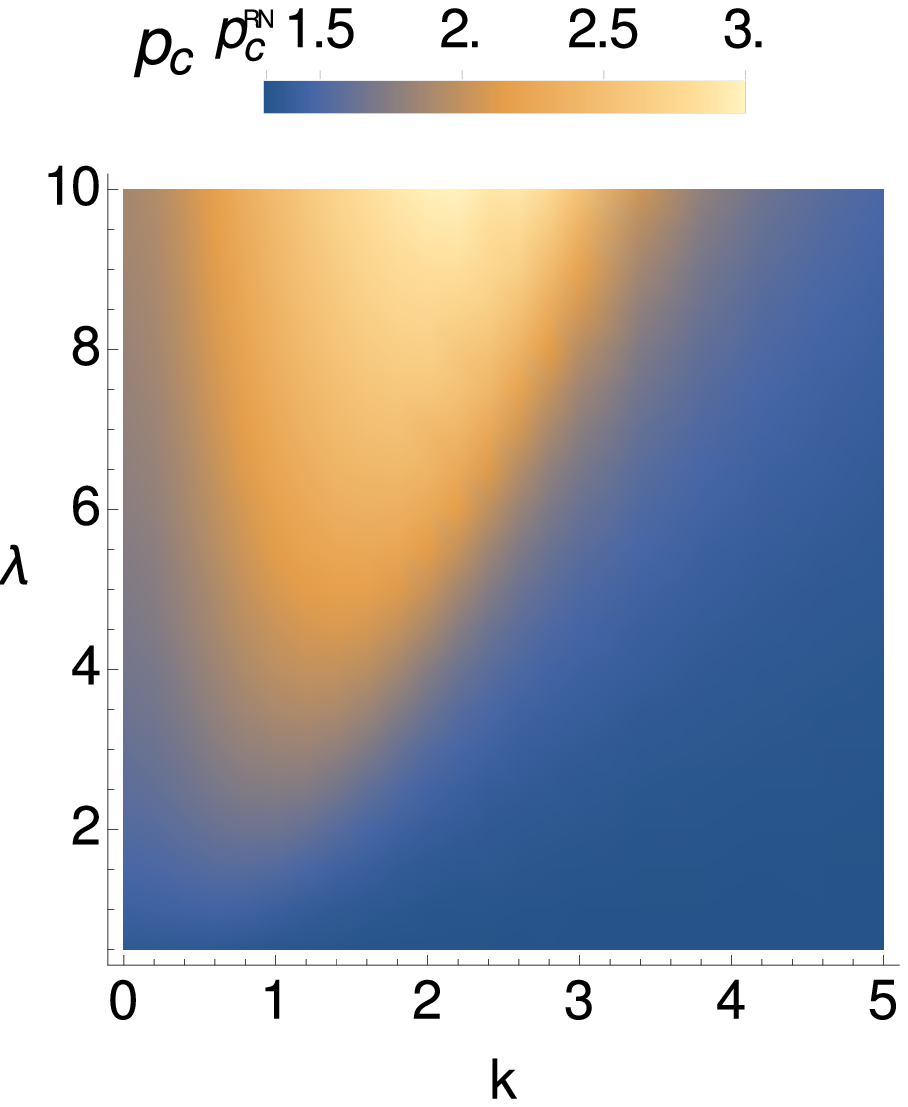}
}
\quad
\subfloat[][\label{Q_sigmaDC_plot} Phase diagram for metal/insulator phases. ($T\sim 10^{-4}$ at large $k$ the accuracy is insufficient)]{
  \includegraphics[width=0.3\linewidth]{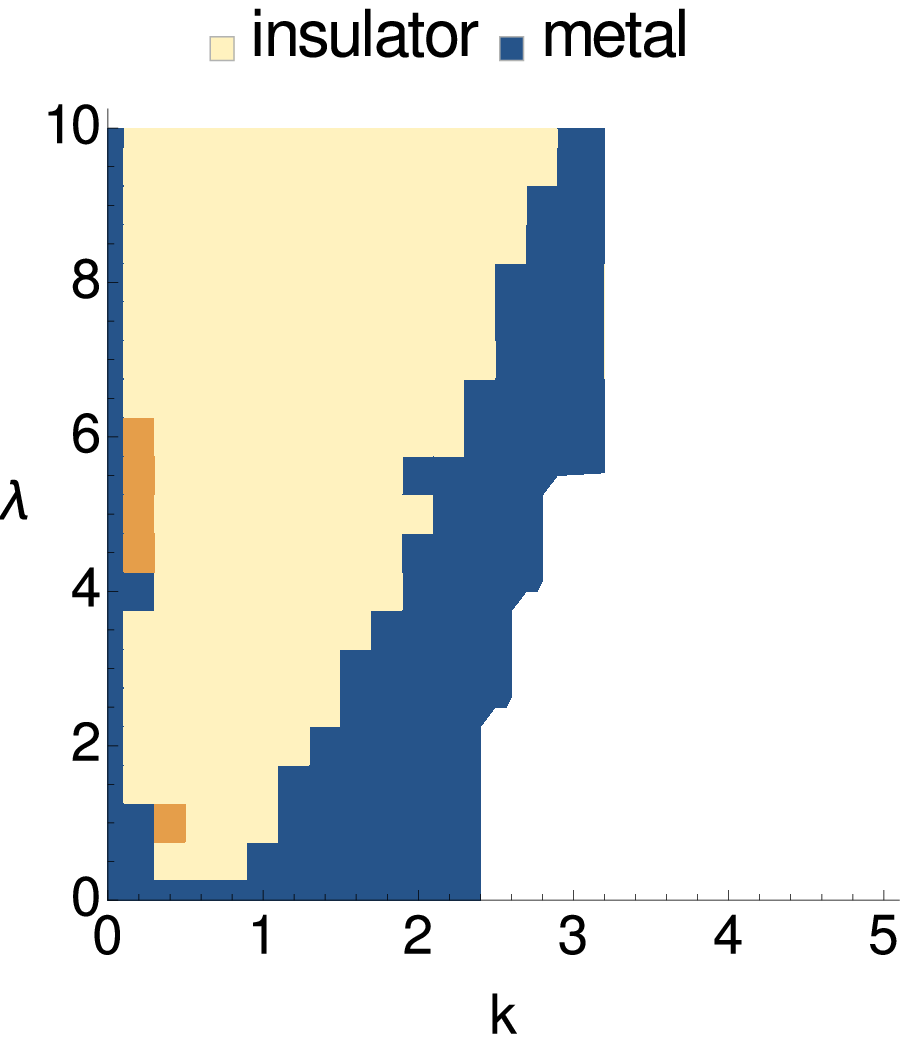}
}
\quad
\subfloat[][\label{Q_chi_plot} Horizon value of the lattice profile $\chi (r_h)$ at low temperature.]{
  \includegraphics[width=0.3\linewidth]{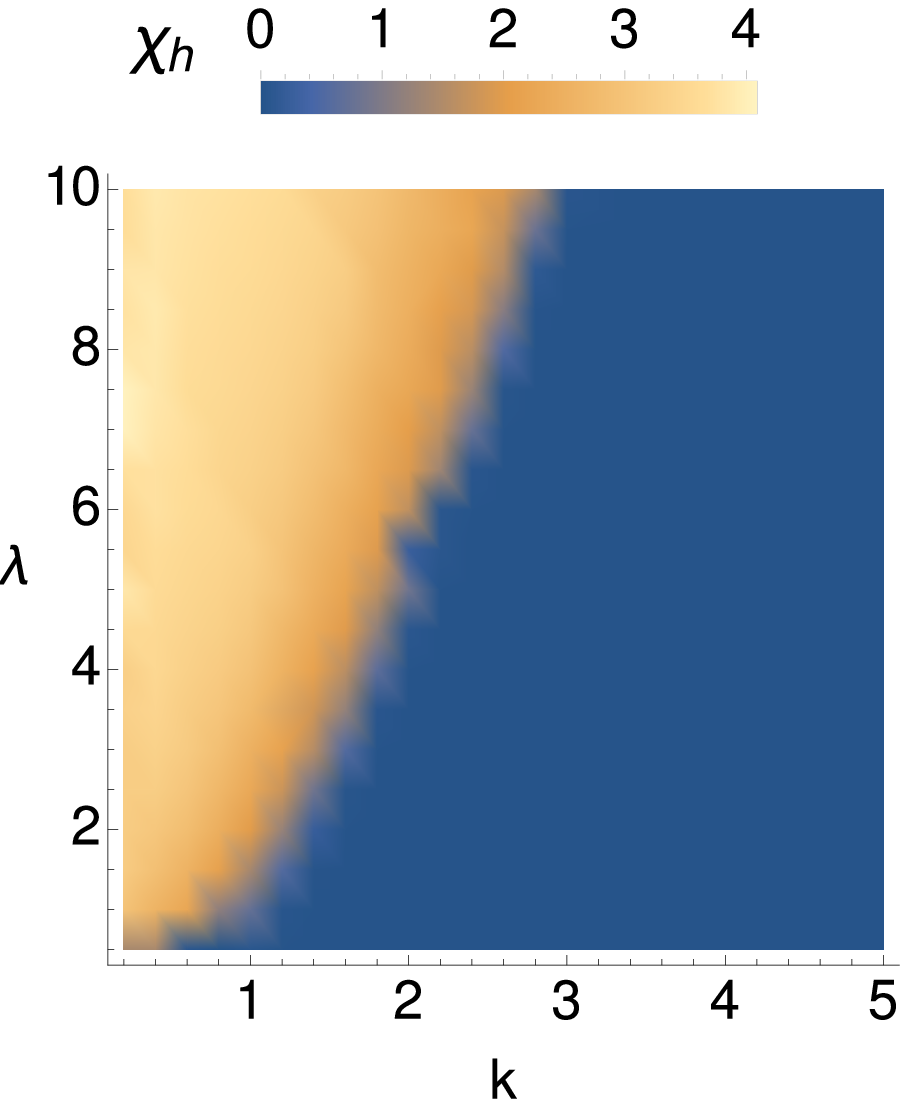}
}
\caption{\label{All_Q_pc_3D} IR physics of the Q-lattice background.}
\end{figure}

For a given set of parameters one needs to solve the nonlinear system of ODEs numerically. This can be done either by a shooting method \cite{Donos:2013eha,Ling:2014laa} or by a relaxation method, which we describe in Appendix \ref{app:relax}. We use both of them independently and cross check the results, which coincide within our numerical accuracy. We scan the space of parameters in the ranges 
$\lambda \in [0.5,10], k \in [0,5]$ and $T \gtrsim 10^{-4}$. Following \cite{Donos:2014uba}, the DC conductivity can be expressed in terms of the horizon data as
\begin{equation}\label{sigmaDC Qlattice}
	\sigma_{DC} = e^{- v_{1h}+2 v_{2h}} \left(1  + \frac{e^{2 v_{1h}} E_h^2 }{2 k^2 \chi_h^2} \right) 
\end{equation}
The phenomenological definition of metals and insulators is based on the behavior of $\sigma_{DC}$ at low temperature 
\cite{Donos:2012js, Donos:2014uba}. In metals 
$\sigma_{DC}$ diverges while in insulators it goes to zero. Hence a reliable test for whether the background is metallic 
or insulating is to measure the sign of the exponent $\sigma_{DC}(T) = T^{\zeta}$, i.e. the derivative in the Log-Log plot, at low temperature.
As in the four-dimensional case, we find both metallic and insulating solutions by evaluating \eqref{sigmaDC Qlattice} 
at low temperatures down to $T \sim 10^{-4}$, see Fig.\,\ref{Q_sigmaDC_plot} for the phase diagram (a similar phase 
diagram was obtained in the four dimensional Q-lattice in \cite{Ling:2015dma}). 

It is pertinent to contrast this characterization with the one obtained in field theory using the memory matrix formalism 
\cite{Hartnoll:2012rj}, in which momentum relaxing configurations are seen to be metallic whenever the operator that breaks translational invariance 
becomes irrelevant in the IR, while they turn out to be insulating if the operator is relevant. 
We can obtain a rough estimate on the relevance of the lattice in the IR by simply plotting the value of $\chi$ at the horizon, 
see Fig.\,\ref{Q_chi_plot}. 
In agreement with the results of \cite{Hartnoll:2012rj}, we note that both criteria of characterizing the metal/insulator transition qualitatively 
coincide in our setup. 
In order to understand the difference between the plots Fig.\,\ref{Q_sigmaDC_plot} and Fig.\,\ref{Q_chi_plot} at zero $k$, 
we recall that even though it appears like the lattice operator is relevant in the IR, due to $k=0$ it does not relax momentum 
any more and the state is metallic.

\begin{figure}[ht]
\centering
 \includegraphics[width=1.\linewidth]{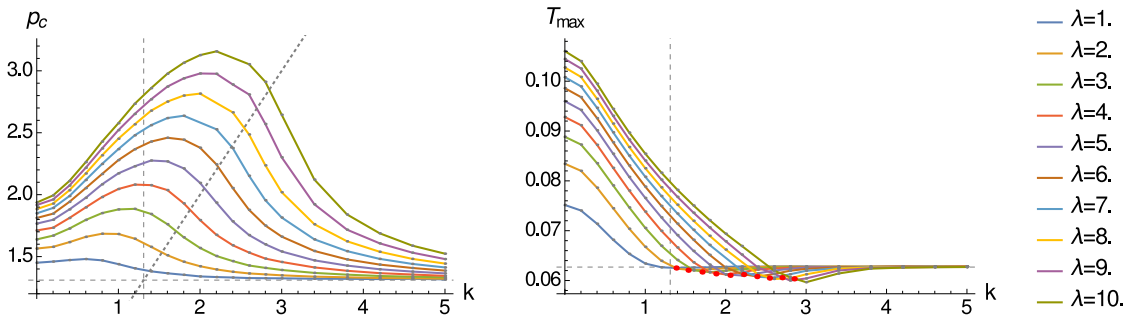}
  \caption{\label{Q_results} The position of the top of the bell curves in various Q-lattice backgrounds. The diagonal line 
  on the $p_c$ plot has unit slope $p_c(k)=k$. The dashed vertical line shows the commensurate value $k=p_c^{RN}$. The horizontal 
  line on the right panel show the value of $T_{max}$ for RN and the red dots mark the points of resonant intersections $p_c(k) = k$.}
\end{figure}

We now proceed to determine the onset of the spatially modulated instabilities in the Q-lattice background in the parameter 
space spanned by $\lambda$ and $k$, expressed in units of $\mu$. One can once again check that a consistent set of perturbations can be chosen to be those of 
RN \eqref{spont_ansatz}, \eqref{delta ds2 RN}, which are governed by equations 
\eqref{lin eom qlattice 1}, \eqref{lin eom qlattice 2}, subject to the boundary conditions \eqref{bc H RN}, \eqref{bc dM RN}. 
In close parallel with  the axion model of Section \ref{sec:axion}, we find that for all studied points in parameter space $(\lambda, k)$ the unstable modes arrange themselves in bell-shaped curves. The coordinates of the tips of the curves, $p_c$ and $T_{max}$, depend on the background parameters, see Fig. \ref{Q_results}.
We do not observe any salient phenomenon when $k=p_c^{RN}$, which means that we do not detect commensurability effects in the Q-lattice. The red dots on the $T_{max}(k)$ plot, the red panel in Fig. \ref{Q_results}, mark the points were the corresponding $p_{c}(T)$ curve crosses the $p_c=k$ line. As we discussed in the Introduction, an enhancement of the critical temperature at these points would show the interference between the helix and the background $p_c(k) = k$. 
The $T_{max}(k)$ plots do not seem to have peaks in these points. Rather, it appears that the red dots are located 
at the minima of the curves. Upon closer inspection we find that this is not the case, so we conclude that 
the interference, either constructive or destructive, is not present.

However, there is an interesting tendency in the dependence of the critical pitch $p_c$ on the lattice momentum $k$. It appears that when the background lattice is strong in the IR, this is to say, for large $\lambda$ in the insulating phase, the dependence tends to be linear with unit slope, i.e. $p_c \approx p_0 + k$ (where the offset $p_0$ is of order of the $p_c^{RN}$). This relation between the momenta is lost when $k$ is large and the background becomes metallic. However, we should stress that this behavior does not possess any resonant or commensurate signature, i.e. $\frac{p_c}{k} \notin \mathbb{N}$.

Interestingly, we observe that for all backgrounds in which the effect of the lattice is small in the IR -- this is, not only solutions for which $\lambda$ is small but also for all metals -- the location of $p_c$ and $T_{max}$ coincides with that of the RN solution, see Fig.
\ref{All_Q_pc_3D}. Moreover, we have 
also verified that for solutions in the metallic regime the shapes of the bell-curves on the $(p, T)$ plane are very close to the translationally invariant solution. 
As mentioned above, based on the results of \cite{Nakamura:2009tf} we expect the formation of the spatially modulated 
instabilities to be an IR phenomenon and our results confirm this expectation since solutions with similar IR geometries 
develop similar instabilities. Unfortunately, because the Q-lattice background is not known analytically, we have not been able 
to make this statement as precise as in \cite{Nakamura:2009tf}. We leave this interesting question for future investigation. \\

\section{\label{sec:helix}Helical background}

The helical background was proposed as a holographic model in \cite{Donos:2012js}. The key ingredient is a (massive) 
vector field $B$, which has a finite helical source on the boundary such that 
 \begin{equation}
 \label{source_B}
 r\rar \infty: \qquad B= \lambda \omega^{(k)}_2
 \end{equation}
 with the helical form defined as in \eqref{helical_forms}. This source breaks the translation symmetry explicitly and defines the pitch $k$ of the background helical structure. 
 In order to study the formation of the spontaneous helix on this background we consider the Lagrangian
  \begin{gather}
 \label{action_helix}
 \mathcal{L} = \mathcal{L}_0 + \mathcal{L}_B + \mathcal{L}_{CS} \\
   \mathcal{L}_B = - \sqrt{-g}  \left( \frac{1}{4}W_{IJ}W^{IJ} + \frac{m^2_B}{2}B_I B^I \right),
 \end{gather}
\noindent where $\mathcal{L}_0$ and $\mathcal{L}_{CS}$ are defined in \eqref{LRN},\eqref{LCS}, $W= dB$ and $m_B$ is the mass of the 
vector field. 
We have modified the Lagrangian of \cite{Donos:2012js} by including the CS term, which triggers the spatially modulated instability, 
and discarded a term of the form $\sim B \wedge F \wedge W$ in order to avoid the direct coupling between the $A$ and $B$ fields and focus 
on the interaction of $A$ with gravity. 
We also set $m_B=0$ for simplicity. 
All of these modifications do not affect the key features of the helical background \cite{Donos:2012js}, namely, 
finite resistivity and a metal-insulator transition.
 
 The distinctive feature of the source (\ref{source_B}) is that it gives rise to a geometry with Bianchi~VII$_0$ symmetry. While breaking the translation and rotation symmetry, this geometry preserves a linear combination of these two. This feature leads to the fact that by choosing a suitable ansatz for the vector and metric fields 
 \begin{gather}
  \label{helical_background}
A = a(r) dt, \qquad B = w(r) \omega^{(k)}_2, \\
\notag
ds^2 = -U(r) dt^2 + \frac{dr^2}{U(r)} + e^{2 v_1 (r)}\big(\omega^{(k)}_1 \big)^2 + e^{2 v_2(r)} \big(\omega^{(k)}_2 \big)^2 + e^{2 v_3(r)}\big(\omega^{(k)}_3 \big)^2,
 \end{gather}
\noindent one can recast the equations of motion as a set of ordinary differential equations for 6~unknown functions $\{U(r), a(r), w(r), v_i(r)\}$, which depend only on the radial coordinate (see \cite{Donos:2012js, Erdmenger:2015qqa}). We note though, that unlike the Q-lattice \eqref{metric Q-lattice} the background (\ref{helical_background}) is explicitly dependent on the $x$ coordinate due to the $\omega$-forms (\ref{helical_forms}). The details of the asymptotic behavior of the ansatz fields as well as the counting of free parameters of the background can be found in \cite{Donos:2012js, Erdmenger:2015qqa}. The bottom line is that black hole solutions are characterized by three independent parameters which we 
take to be $T/\mu$, $\lambda/\mu$ and $k/\mu$. 

To proceed with the study of spatially modulated instabilities we must adopt an ansatz for the fluctuations of the components of the gauge field. If one tries to introduce an ansatz similar to (\ref{spont_ansatz}), one observes that when $p \neq k$ the helical form $\omega^{(p)}_2$ does not respect the Bianchi VII$_{0}$ symmetry of the background. When expressed in terms of the background helical forms $\omega^{(k)}$ it includes two modes which propagate differently in the bulk.
\begin{equation}
\label{expand_form}
 \omega_2^{(p)} = \cos \! \Big((k-p)x \Big) \, \omega_2^{(k)} + \sin \! \Big((k-p)x \Big) \, \omega_3^{(k)}.
\end{equation}
Therefore this ansatz does not provide any simplification to the problem. Hence one can equally well use a more general ansatz, which includes two arbitrary functions with arbitrary $x$-dependence 
\begin{align}
\label{helical_ansatz}
 \delta A = \delta A_2(r,x) \omega^{(k)}_2 + \delta A_3(r,x) \omega^{(k)}_3 . 
\end{align}
On top of that one must turn on the fluctuations of the components of the metric and vector field $B$, which couple linearly to $\delta A$:
\begin{gather}
\label{helical pert on the helix}
\delta B = \delta B_t(r,x) dt \\
\delta (ds^2) = 2 \delta Q_2(r,x) e^{2 v_2} \omega^{(k)}_2 dt + 2 \delta Q_3 (r,x) e^{2 v_3} \omega^{(k)}_3  dt. 
\end{gather}

The linearized equations of motion for this ansatz can be brought into a form which does not  possess any explicit dependence of $x$. 
This allows us to make a Fourier transform along $x$-coordinate and consider the modes\footnote{The time component 
of $\delta B$ has the same phase as $\delta A_2$ and $\delta Q_2$ because due to nonzero background $B \sim \omega_2$ it 
interacts in kinetic term with $\delta Q_2$ and $\p_x \delta Q_3$. } 
\begin{align}
\label{modes}
 \delta A_2(r,x) &= \cos(q x) \delta A_2(r), &  \delta A_3(r,x) &= \sin(q x) \delta A_3(r), \\
 \notag
 \delta Q_2(r,x) &= \cos(q x) \delta Q_2(r), &  \delta Q_3(r,x) &= \sin(q x) Q_3(r), \\
 \notag
 \delta B_t (r,x) & = \cos(q x) \delta B_t (r).
\end{align}
The equations of motion are reduced then to the system of ODE's \eqref{dEOMs}. The physical meaning of the Fourier parameter $q$ is clear once one recalls the expansion of the $\omega^{(p)}$ form (\ref{expand_form}). The helical instability with pitch $p$ corresponds to the solution with $\delta A_2 = \delta A_3$ and $q=k-p$.

In order to produce the bell curves we look for the values of $q$ at which nontrivial solutions to the system exist with boundary 
conditions which correspond to setting the dual sources to zero. This means that at the $AdS$ boundary $r\rar\infty$ the modes behave as 
\begin{align}
r\rar \infty:\qquad \delta A_i(r) &= \frac{A_i^{(2)}}{r^2} + \dots, & \delta Q_i(r) &= \frac{Q_i^{(2)}}{r^4} + \dots, & \delta B_t(r)  &= \frac{B_t^{(2)}}{r^2}+\dots.
\end{align}
At the horizon the boundary conditions are set by regularity
\begin{align*}
r&\rar r_h: & \delta A_i(r) &= A_i^{(h)}+\dots, & \delta B_t(r) &= B_t^{(h)}(r-r_h)+\dots, & \delta Q_i(r) &= Q_i^{(h)}(r-r_h)+\dots. 
\end{align*}

It is useful to note that the ansatz (\ref{modes}) does not change under the operation 
\begin{equation}
\label{pairs}
q \rar -q, \qquad \delta A_3 \rar - \delta A_3, \qquad \delta Q_3 \rar -\delta Q_3, 
\end{equation}
henceforth all nontrivial solutions come in degenerate pairs. However, this symmetry is just an artifact and each solution in the pair describes the same function $\delta A_3(r,x)$. 

We span the parameter space of the backgrounds by picking particular values of $\lambda$ and $k$. For each set the study of the eigenvalues of $q$ at various temperatures results in the bell curve similar to Fig.\,\ref{bell_RN}. Similarly to the case of Q-lattice we then trace the position of the tip of the bell depending on $\lambda$ and $k$.

Once again the background is obtained numerically by solving nonlinear ODE's. In the earlier works \cite{Donos:2012js, Erdmenger:2015qqa} 
this was done by a shooting method, but we found that a finite difference relaxation method is much more efficient, see Appendix \ref{app:relax}. The main reason is that this method does not suffer from the presence of divergent modes in the spectrum, which 
occur because of the singularity in the equations at the boundaries \cite{press1992numerical}. Thus, it is much more robust. 
Once the background solution is obtained as a set of discrete function values on the grid, the Sturm-Liouville problem for the fluctuations can be solved on the grid as well. We check the reliability of our numerical results by comparing the values obtained by the relaxation method and the conventional shooting method for a set of points in parameter space finding agreement within numerical error bars. 

\begin{figure}[ht]
\centering
 \includegraphics[width=1.\linewidth]{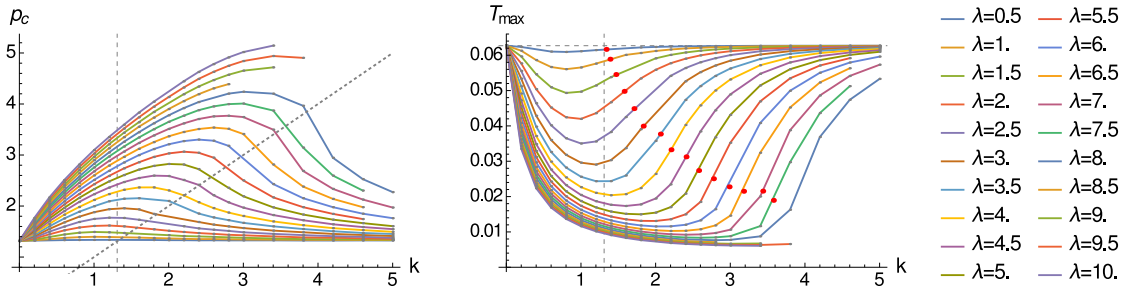}
  \caption{\label{helix_results} The position of the top of the bell curves in various helical backgrounds. The dashed line on the  $p_c(k)$ plot shows the commensurate value $k=p_c^{RN}$ \eqref{RNvalues}. The dotted line has unit slope $p_{c}=k$. The red dots on the $T_{max}(k)$ plots show the points of resonant intersections $p_c(k) = k$. The dashed line on $T_{max}(k)$ plot show the value of $T_{max}^{RN}$.}
\end{figure}

The results which we get for the helical background are shown on Fig.\,\ref{helix_results}. We span the same parameter space as in the previously considered case of Q-lattice. At large $\lambda$ and large $k$ the procedure calls for higher precision than the one we 
have reached, so we cut the curves at some value of $k$ since this does not alter our main results. 
Unfortunately, even in this most sophisticated example we  do not see any features at the commensurability point $k=p_c^{RN}$. 
Looking for indications of interference  we determine the background parameters for which the $p_c(k)$ curves cross the $p_c=k$ line and check that at the corresponding points on the temperature plots (marked in red) there are no features either. It is worth mentioning that in the Bianchi VII background we observe a suppression of the transition temperature similar to the case of linear axions. Once again, we 
observe that at large $\lambda$ in the insulating regime the dependence of $p_c$ with $k$ saturates at the line with unit slope 
$p_c(k) \approx p_0 + k$, where $p_0 \sim p_c^{RN}$.

\begin{figure}[ht]
\subfloat[][\label{3DQ_results} Dependence of $p_c$ on the background parameters.]{
  \includegraphics[width=0.3\linewidth]{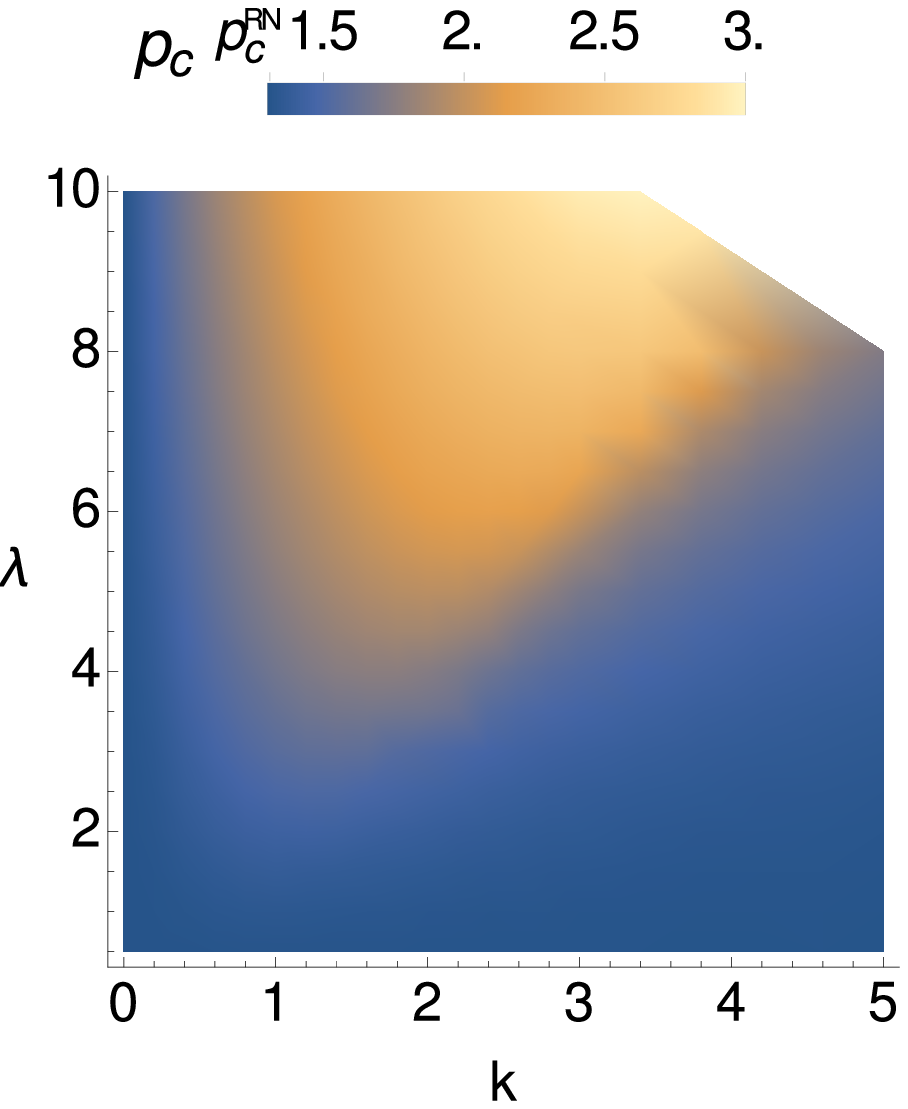}
}
\quad
\subfloat[][\label{helix_sigmaDC_plot} Phase diagram for metal/insulator transition.(Orange region-- insufficient data.)]{
  \includegraphics[width=0.3\linewidth]{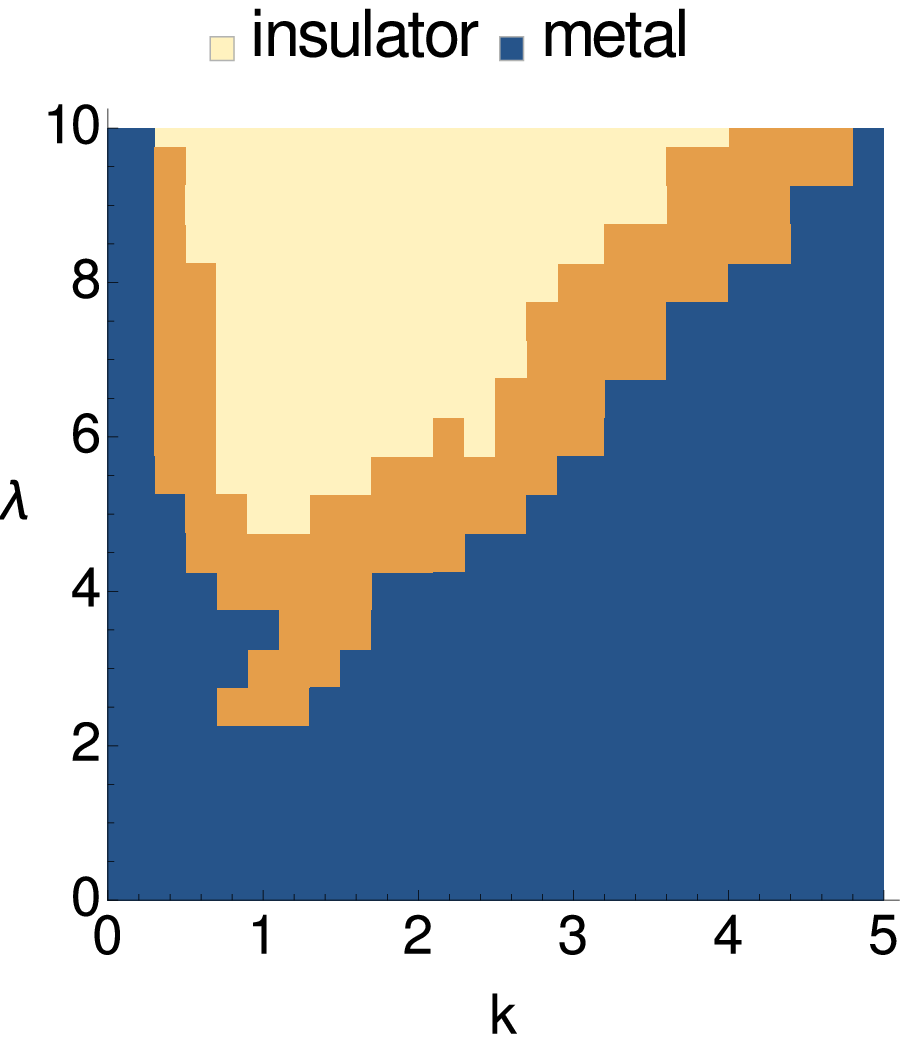}
}
\quad
\subfloat[][\label{helix_ws_plot} Horizon value of the lattice profile $w(r_h)$ at low temperature $T\sim 10^{-5}$]{
  \includegraphics[width=0.3\linewidth]{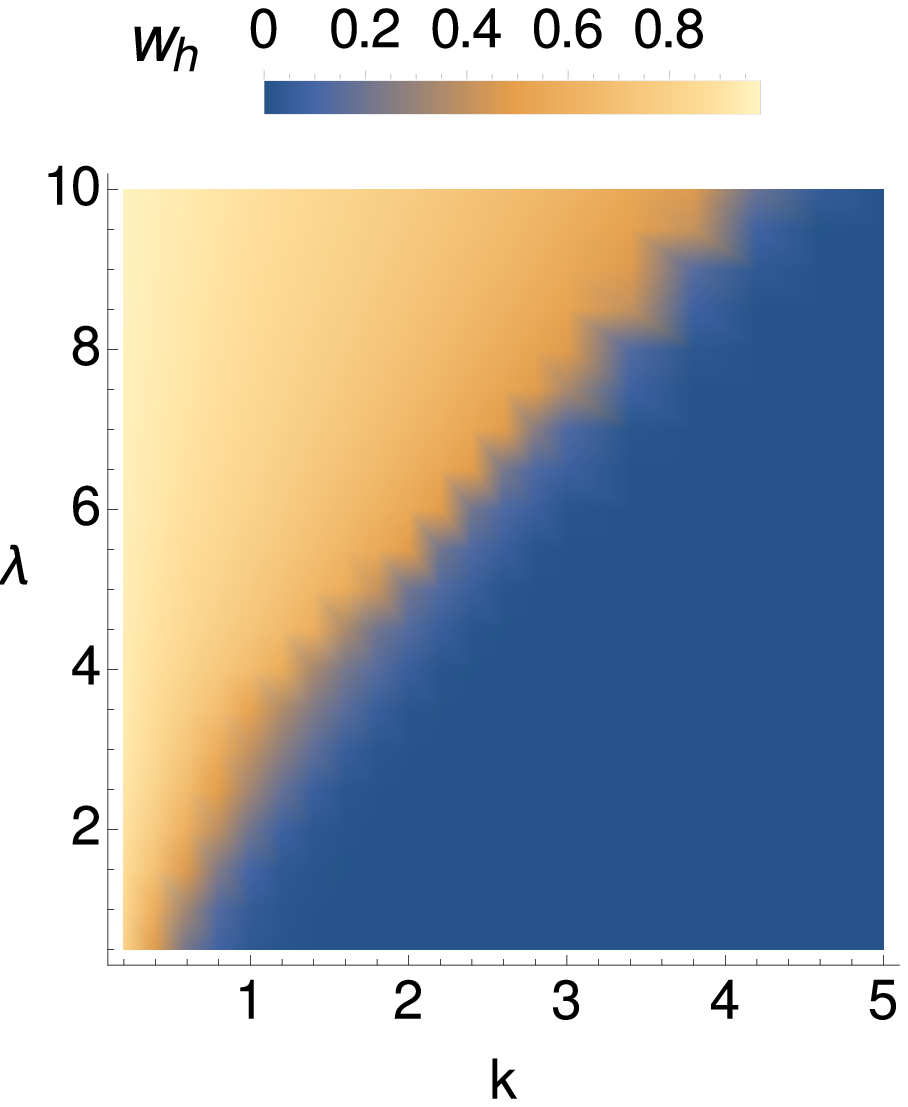}
}
\caption{IR physics of helical background. }

\end{figure}

As in the case of the Q-lattice, for large $k$ the curves converge to the values in the RN background (\ref{RNvalues}), so 
we expect the corresponding backgrounds to be metallic. 
In order to check this we determine the metal/insulator phase diagram using the formula 
\begin{equation}\label{sigmaDC helix}
 	\sigma_{DC}= \frac{1}{2} e^{ - v_1 + v_2 + v_3 } \left( 1 +  
	\frac{e^{2(v_1 + v_2 + v_3)} a'^2}{k^2  [ (e^{2 v_2} - e^{2 v_3})^2 + e^{2 v_2} w^2 ] }  \right) \bigg | _{r = r_h}  
\end{equation}
\noindent 
which we derive in Appendix \ref{app:sigmaDC} following an argument in \cite{Donos:2014uba}. We evaluate \eqref{sigmaDC helix}  
at temperatures as low as $T=10^{-5}$, finding the diagram on Fig.\,\ref{helix_sigmaDC_plot}.
Indeed, in close analogy with section \ref{sec:Qlattice}, the study of the lattice profile $w$ at the horizon shows that the metallic phase corresponds to vanishing $w(r_h)$ (see Fig.\,\ref{helix_ws_plot}) and vice versa: nonzero $w(r_h)$ at finite $k$ leads to 
the insulating behavior. 
In the same fashion as in the previous examples, the nontrivial relation between $p_c$ and $k$ is observed only in the insulating region (see Fig.\,\ref{3DQ_results}).This confirms our hypothesis that the spontaneous helix is indeed sensitive only to the IR geometry, 
like the DC conductivity.

\section{\label{Conclusion}Conclusions}

In this work we have studied the spontaneous formation of a helical current state on homogeneous holographic lattices.
We have focused on the onset of such instability, so that a perturbative analysis suffices to extract the physics of interest.
For all the gravitational backgrounds under consideration, we find that 
it is possible to characterize the onset of the instabilities by only two parameters: the critical temperature $T_{max}$ -- the maximum 
temperature at which the condensate develops --  and the critical momentum $p_c$ -- the momentum singled out by the unstable 
mode at this temperature scale. Despite the differences in the considered backgrounds, we find that the phenomenology of all three examples is quite similar.

Our results show that, as in the translational invariant case, the spontaneous formation of the helical phase is an IR effect, 
or in other words, that the unstable modes localize near the horizon. In particular, this means that for metals, i.e. configurations
for which the lattice is irrelevant in the IR, the structure of the instabilities can be well approximated by the translational invariant results.
On the other hand, we find a richer structure in insulating backgrounds since in this case the lattice remains relevant in the near horizon. 
In this regard, our results are analogous to those of \cite{Erdmenger:2015qqa, Ling:2015dma}, where it was observed that the physics of 
holographic superconductors is affected by the underlying lattice only when the normal state is insulating.

It is thus in the insulating phase when it is meaningful to ask if there is any interesting correlation between the lattice parameters 
and the ones that characterize the spontaneously generated helix.   
In all three examples, which we consider, we do not observe any commensurability phenomena, which one might have expected to arise due to the interplay of the period of the spontaneous helix and the lattice spacing parameters. 
%
Instead, in all the studied cases we observe a generic linear relation of the form 
$p_c(k) = k + p_0$, where $k$ is the momentum scale of the background and the offset $p_0$ is close to 
the natural momentum of the helix. This behavior is not expected for two interacting periodic systems,
which would rather tend to have equal wave vectors without any offset. 

The absence of commensurability effects is not entirely surprising 
if we look more closely at the specific features of the solutions that we are studying. 
Indeed, the common ingredient of the linear axion, Q-lattice and helical geometries is that the product of spatial translations and the translations along a certain global symmetry group can be broken down to the diagonal subgroup. Therefore, these configurations do break translational invariance, but the aforementioned diagonal combination remains as a continuous symmetry of the theory. 
This is what allows us to construct the backgrounds by solving ODE's in the first place, but it also suggests that there is no 
physical periodic structure in the model.  
Therefore the homogeneous lattice backgrounds should be naturally considered as obeying some kind of \textit{linear}, rather then oscillatory, dynamics. This point of view also helps to explain the common linear behavior of $p_c (k)$ curves in all our examples. 
A simple mechanical analogy to the homogeneous lattice is not an oscillator, but rather a brick which slides with constant speed $v$. The spontaneous helix in this analogy would be another brick, which slides on top of the first one due to an applied force, with the speed $u_0$ stabilized by friction. As a result, the total speed of the second brick is the sum of these two: $u = v + u_0$, exactly the linear law which we observe on our plots. 

We would like to note here that in this study we have imposed restrictive constraints on the possible interactions of our probe with the background fields. One can imagine including the specific interactions or the specific probes, which would mimic the commensurability effects even in the homogeneous lattice backgrounds. But these effects would be very much dependent on the model under consideration. What we show in the present study, is that the homogeneous lattices do not describe the periodic structures \textit{generically}.

Summing up, our work uncovers some generic features of spontaneous symmetry breaking in the presence of 
holographic homogeneous lattice. By pointing out that commensurability effects are not properly accounted for, we hope to help delimit the applicability of these otherwise powerful theoretical tools.

\acknowledgments

We thank Jan Zaanen and Koenraad Schalm for valuable comments. 
T.A. was supported by the European Research Council under the European Union's Seventh Framework Programme
(ERC Grant agreement 307955). He also thanks the Niels Bohr Institute, NORDITA and the Institute of Physics at 
University of Amsterdam for their hospitality during the completion of this work.
A.K. is supported by a VICI grant of the Netherlands Organization for Scientific Research (NWO), by the Netherlands Organization for Scientific Research/Ministry of Science and Education (NWO/OCW) and by the Foundation for Research into Fundamental Matter (FOM). The work of A.K. is partially supported by RFBR grant 	15-02-02092a.
A.K. thanks Rudolf Peierls Centre for Theoretical Physics in the University of Oxford for warm hospitality during the completion of this project.
We acknowledge the partial support of the INFN and the hospitality of the Galileo Galilei Institute for Theoretical Physics where this 
work was initiated.

\appendix

\section{\label{app:eoms}Equations of motion}

In this appendix we collect the explicit expressions for various equations of motion referred to in the main text.

\subsection*{RN and linear axion perturbations}

\begin{align}
\label{lin eom axion 1}
	r^2 \qf'' + r \qf' -  \frac{(p^2 + \alpha^2 + 4 U)}{U}  \qf +r^2 a' \af '   &= 0 \\  
\label{lin eom axion 2}
	r^2 \af'' + r^2 \left( \frac{U'}{U} + \frac{1}{r} \right) \af' + \left( \gamma p  r a' - p^2 \right) \frac{\af}{U} - 2 \frac{r a'}{U} \qf &= 0 
\end{align}
The perturbation equations around the RN solution are obtained by letting $\alpha = 0$ in  
\eqref{lin eom axion 1}, \eqref{lin eom axion 2} above.

\subsection*{Q-lattice background}

\begin{align}
\label{bck eom qlattice 1}
	\chi'' + \left( \frac{U'}{U} + v_1' + v_2' \right) \chi' - \frac{(m^2 + e^{2 v_1} k^2)}{U} \chi &= 0 \\
	a'' + a' (v_1' + 2 v_2') &= 0 \\
	U' \left( \frac{v_1'}{2} + v_2' \right) + U ( 2 v_1' v_2' + v_2'^2 ) - \frac{1}{2} U \chi'^2 + \frac{1}{2} ( e^{- 2 v_1} k^2 + m^2 ) \chi^2 
	+ \frac{1}{4} a'^2 - 6 &= 0   \\
v_1'' +  v_1'^2 + 2 v_1' v_2' + \frac{U' v_1'}{U} + \frac{1}{U}\left( e^{- 2 v_1} k^2 + \frac{m^2}{3} \right) \chi^2 +\frac{1}{U} \left( \frac{1}{6} a'^2 - 4 \right) &= 0 \\
\label{bck eom qlattice 5}
	v_2'' + 2 v_2'^2 + v_1' v_2' + \frac{U' v_2'}{U} +\frac{m^2}{3 U}  \chi^2 +\frac{1}{U} \left( \frac{1}{6} a'^2 - 4 \right) &= 0 
\end{align}

\subsection*{Q-lattice perturbations}

\begin{align}
\label{lin eom qlattice 1}
	U \af'' + (  U' + U v_1' ) \af' + e^{- v_1} (\gamma p  a'  - p^2 e^{- v_1}   ) \af + a' \qf'  - 2 a' v_2' \qf &= 0 \\
\label{lin eom qlattice 2}
	\qf'' + v_1' \qf' + ( 2 m^2 \chi^2 + a'^2 + 6 U' v_2' - 3 p^2 e^{- 2 v_1}  - 24  ) \frac{\qf}{3 U} + a' \af' &= 0 
\end{align}

\subsection*{Helix perturbation}

\begin{align}
\label{dEOMs}
U \delta A_2''  + & \left[U' + U(v_1' - v_2'+ v_3') \right]  \delta A_2' + e^{2 v_2} a' Q_2' \\
\notag
& -  e^{-2(v_1 + v_3)} \left[ e^{2 v_2} k^2 + e^{2 v_3} q^2 - \gamma e^{v_1 + v_2 + v_3} k a'  \right] \delta A_2  \\
\notag
& - e^{-2(v_1 + v_3)} \left[  (e^{2 v_2} + e^{2 v_3}) k q - \gamma e^{v_1 + v_2 + v_3} q a' \right] \delta A_3   = 0 \\
\notag
U \delta A_3''  + &\left[U' + U(v_1' + v_2' - v_3') \right]  \delta A_3' + e^{2 v_3} a' Q_3' \\
\notag
& -  e^{-2(v_1 + v_2)} \left[ e^{2 v_3} k^2 + e^{2 v_2} q^2 - \gamma e^{v_1 + v_2 + v_3} k a'  \right] \delta A_3  \\
\notag
& - e^{-2(v_1 + v_2)} \left[  (e^{2 v_2} + e^{2 v_3}) k q - \gamma e^{v_1 + v_2 + v_3} q a' \right] \delta A_2   = 0 \\
\notag
U Q_2'' + & U \left[v_1' + 3 v_2' + v_3'\right] Q_2' + e^{-2 v_2} \left[U a' \delta A_2' + U w' \delta B_t' \right]\\
\notag
- & e^{-2(v_1+v_2)} \left[ e^{2 v_3} k^2 + e^{2 v_2} q^2 + e^{2 v_1} U w'^2 \right] Q_2 - e^{-2(v_1 + v_2)}\left[e^{2 v_2} + e^{2 v_3} \right] k q Q_3 =0 \\
\notag
U Q_3'' + & U \left[v_1' +  v_2' + 3 v_3'\right] Q_3' + e^{-2 v_3} \left[U a' \delta A_3' - k q e^{-2v_1} w \delta B_t \right]\\
\notag
- & e^{-2(v_1+v_3)} \left[ e^{2 v_2} k^2 + e^{2 v_3} q^2 + k^2 w^2 \right] Q_3 - e^{-2(v_1 + v_3)}\left[e^{2 v_2} + e^{2 v_3} \right] k q Q_2 =0 \\
\notag
U \delta B_t'' + & U \left[ v_1' + v_2' + v_3' \right] \delta B_t' - e^{-2 v_1} q^2 \delta B_t - e^{-2 v_1} k q w \, Q_3 \\
\notag
+ & U w' Q_2' + \left[ (U' - 2 U v_2') w' - e^{-2(v_1 - v_2 +v_3)} k^2 w \right] Q_2 = 0.
\end{align}

\section{\label{app:relax}Numerical finite difference method}

In order to obtain the Q-lattice and helical backgrounds we need to solve numerically a boundary value problem for a given set of ODEs. This can be done either by a shooting method or by finite difference derivative method on the grid (i.e. relaxation method) \cite{trefethen2000spectral}. 
In the present paper we used both methods independently and checked that the results coincide in order to ensure the quality of our numerics. The shooting method is described in detail in e.g. \cite{Donos:2012js,Donos:2013eha, Erdmenger:2015qqa}. In this appendix we outline the relaxation 
method which we use. 

Making the coordinate change $\xi = \frac{r_h}{r}$, we reduce the computation region to the interval $\xi \in [0,1]$, where the horizon 
is located at $\xi=1$ and the AdS boundary at $\xi=0$. We introduce a discrete grid with $N$ nodes, including the endpoints of the
interval, and approximate the functions by a set of their values at these points. Solving for $m$ functions we get $mN$ variables $\vec{f}$. We also calculate the values of the equations at each node by approximating the derivatives with finite differences. 
On the endpoints, the equations are singular, which is remedied by replacing them by the appropriate boundary conditions. 
At the end of the day we reduce the problem to a system of $mN$ nonlinear algebraic equations $\vec{E}$ on $mN$ variables. This system is solved by means of the Newton-Raphson method: at each step, the Jacobian $J$ of the system, a $mN \times mN$ matrix, is computed and the increment of the variables is obtained as
\begin{equation}
\label{linear_solve}
 \delta \vec{f} = - J^{-1} \vec{E}.
\end{equation}
The procedure is iterated until the values of the equations and boundary conditions become numerically small (down to $10^{-10}$) everywhere on the grid. 

There are several technical features, which we would like to point out:

\begin{itemize}
\item Before performing the numerical calculations we rescale the functions in such a way that they assume finite values at the 
boundaries, i.e. we extract the divergent and vanishing asymptotics.

\item At the horizon we obtain the boundary condition by expanding the singular equations of motion and taking the leading part. This is analogous to series expansion of the functions, used in the shooting method. 

\item In the problem under consideration it is very convenient to have direct control over the parameters $\lambda, k$ and temperature $T$. This means that the value of $r_h$, which enters in the boundary condition after we rescale to unit interval, is not known a priori. 
It is useful to promote the radius of horizon to be an additional function to solve for, i.e. $r_h = r_h(\xi)$. The equation of motion is simply $r_h''(\xi) = 0$ and boundary conditions are $r_h'(0)=r_h'(1)=0$. This allows us to generate the solutions for a given $T$, which facilitates the calculation of the bell curve and also helps to avoid generating the solution for extremal black holes. 

\item We use an equally spaced lattice with $N=100$ nodes and we approximate the derivatives with two nearest neighbor differences. 
Had we used a pseudospectral approximation for derivatives we would get the same precision with fewer nodes in the lattice. However, this approach requires introducing a Chebyshev lattice, which suffers from poor convergence when the solutions are non-analytic functions \cite{trefethen2000spectral}. This is the case with square root behavior of scalar in the Q-lattice and logarithms in the helix, so we avoid using pseudospectral methods.

\item We make use of \texttt{NDSolve'FiniteDifferenceDerivative} procedure in Wolfram Mathematica 10 \cite{mathematica10} in order 
to produce the differentiation matrices and \texttt{LinearSolve} function in order to solve eq.(\ref{linear_solve}). 
\end{itemize}

In the relaxation framework there is an elegant procedure to look for the unstable linear fluctuations. The equations of motion 
for fluctuations $\delta f$ are linear differential equations which depend on the additional parameter $p$ (or $q$) \eqref{lin eom qlattice 1}, \eqref{dEOMs}. Upon discretization on the grid they give rise to the system of linear algebraic equations. These equations always have trivial solution $\delta f = 0$, but at specific values of $p$ the other, nontrivial, solution appears. For a linear system that means that the determinant $\mathcal{D}(p)$ of the matrix of the linear system vanishes. Therefore, the problem of finding the pitch of the unstable mode reduces to 
the search for the zeros of $\mathcal{D}(p)$.

\section{\label{app:sigmaDC}DC conductivity in the helical background}

In this appendix we derive the formula for the DC conductivity of the helical background \eqref{helical_background}, 
given in \eqref{sigmaDC helix} of the main text. Following \cite{Donos:2014uba}, our first step is to write 
down an ansatz for the linear perturbations that contains a constant electric field. We have verified that 
a such a consistent set of perturbations can be taken to be 
\begin{equation}
	\delta ( ds^2)  = e^{2 v_1(r)} ( h_{t x}(r) dt  +  h_{t x}(r) dr  ) \omega_1
\end{equation}
\begin{align}
	\delta A &= - E t + {\cal A} (r) \omega_1 \\
	\delta B &= {\cal B}(r) \omega_3 
\end{align}
A consistent ansatz for the linear perturbations relevant to the computation of the frequency-dependent 
conductivity was previously found in \cite{Erdmenger:2015qqa}. In addition to the terms above they also included a metric 
fluctuation proportional to $\omega_2 \omega_3$, but we have checked that this mode decouples in the zero frequency limit. 

The remaining part of the calculation closely follows \cite{Donos:2014uba}. Before studying the perturbation equations,
we observe that the background equation of motion for $a(r)$ can be solved by
\begin{equation}\label{q def}
	q = e^{v_1 + v_2 + v_3} a' - \frac{1}{2} k \kappa w^2 
\end{equation}
\noindent where $q$ is an integration constant related to the background charge density. 

The Maxwell equation for ${\cal A}$ can be integrated once to give
\begin{equation}\label{j def}
 	j = \left( q + \frac{1}{2} k \kappa w^2 \right) h_{tx} + e^{- v_1 + v_2 + v_3} U {\cal A}'
 \end{equation} 
The $r x$ component of Einstein's equations can be used to solve algebraically for $h_{rx}$. In what follows we shall assume that
we are doing so and replacing $h_{rx}$ in all equations. The $tx$ component of Einstein's equation implies
\begin{align}\label{eq htx}
\nonumber
	h_{tx}'' + (3 v_1' + v_2' + v_3') h_{tx}' - \frac{k^2}{U} e^{-2(v_1 + v_2 + v_3)} [  (e^{2 v_2} - e^{2 v_3})^2 + e^{2 v_2} w^2 ] h_{tx} \\
	+ e^{- (3 v_1 + v_2 + v_3) }\left( q + k \kappa w^2 \right) {\cal A}' = 0 
\end{align}
The equation for ${\cal B}$ is complicated but for 
our purposes it suffices to note that it only involves ${\cal B}$ and $E$ but not any other linear perturbation. Moreover,
note that ${\cal B}$ does not appear in \eqref{q def}, \eqref{j def}, \eqref{eq htx}. Hence, it decouples from the system and 
can be ignored in this analysis. However, we stress that setting ${\cal B} = 0$ yields to an inconsistency. 

Let us now impose the boundary conditions. In the UV, we set the source of the metric perturbation to zero which implies
$h_{t x} \sim r^{-2}$. In addition, we find that the near boundary expansion for ${\cal A}$ is of the form 
\begin{equation}
	{\cal A} = {\cal A}^{(0)} + \frac{1}{r} {\cal A}^{(1)} + \frac{1}{r^2} {\cal A}^{(2)} + \ldots
\end{equation}
Note that at zero frequency $ {\cal A}^{(1)} $ and all possible log terms vanish since they are proportional to $\omega$. 
Recalling that $v_i \sim \log r$ and $U \sim r^2$, we learn from \eqref{j def} that 
\begin{equation}\label{j and A2}
	j = - 2 {\cal A}^{(2)}
\end{equation}
Our goal is now to evaluate \eqref{j def} at the horizon, located at $r = r_h$. To do so, we recall that regularity of ${\cal A}$ implies that 
near the horizon we must have
\begin{equation}\label{A hor}
	{\cal A} = - \frac{E}{4 \pi T} \log ( r- r_h )  + \ldots , \qquad  \Rightarrow  \qquad {\cal A'} = - \frac{E}{U} + \ldots
\end{equation}
Using this in \eqref{eq htx} we learn that
\begin{equation}\label{htx hor}
	h_{tx} (r_h) = - \frac{e^{- v_1 + v_2 + v_3} (2 q + k \kappa w^2) E }{2 k^2   [  (e^{2 v_2} - e^{2 v_3})^2 + e^{2 v_2} w^2 ]  } \bigg | _{r = r_h}
	+ \ldots
\end{equation}
Using \eqref{A hor} and \eqref{htx hor} we can evaluate \eqref{j def} at the horizon arriving at our desired 
formula for the DC conductivity
\begin{align}
	\sigma_{DC} = - \frac{j}{2E}  
%
%
\label{sigma app}
	&= \frac{1}{2} e^{ - v_1 + v_2 + v_3 } \left( 1 +  
	\frac{e^{2(v_1 + v_2 + v_3)} a'^2}{k^2  [ (e^{2 v_2} - e^{2 v_3})^2 + e^{2 v_2} w^2 ] }  \right) \bigg | _{r = r_H}  
\end{align}
The overall factor of $-1/2$ has been chosen so that, at non-zero frequency, the conductivity is defined as the ratio 
\begin{equation}\label{sigma AC helix}
	\sigma = \frac{{\cal A}^{(2)}}{i \omega {\cal A}^{(0)}}
\end{equation}
Both the overall factor and possible contact terms in \eqref{sigma AC helix} can be obtained by 
carrying out the holographic renormalization 
procedure. They are not important to us since we are only interested in the temperature dependence of $\sigma_{DC}$.   
We finish this appendix by mentioning that we have checked numerically that \eqref{sigma app} correctly matches the 
small frequency limit of the AC conductivity for several values of the parameters. 

\bibliographystyle{JHEP-2}
\bibliography{stripes_lattice}

\end{document}